\documentstyle[proceedings,psfig]{crckapb}  

\def\figdir{fig}
\def\note #1]{{\bf #1]}}
\def\etal{{\it et al.}}
\def\eg{{\it e.g.}}
\def\cf{{\it cf.}}
\def\ie{{\it i.e.}}
\def\bibref{\par\noindent
	\hangindent=0.7 true cm
	\hangafter=1{}}
\def\dd{{\rm d}}
\def\CG{{\cal G}}
\def\CH{{\cal H}}
\def\CK{{\cal K}}
\def\CHone{\CH_1}
\def\CHtwo{\CH_2}
\def\rt{r_{\rm t}}
\def\rcz{r_{\rm cz}}
\def\pcz{p_{\rm cz}}
\def\pph{p_{\rm ph}}
\def\dcz{d_{\rm cz}}
\def\sad{s_{\rm ad}}
\def\sph{s_{\rm ph}}
\def\pph{p_{\rm ph}}
\def\nablaad{\nabla_{\rm ad}}
\def\Fcon{F_{\rm con}}
\def\Fsurf{F_{\rm surf}}

\def\alphac{\alpha_{\rm c}}
\def\be{\begin{equation}}
\def\ee{\end{equation}}

\def\cm{\,{\rm cm}}
\def\dyn{\,{\rm dyn}}

\def\muHz{\,\mu{\rm Hz}}

\begin{opening}
  \title{Effects of convection \hskip -0.7em \protect \\ 
  on the mean solar structure} 
  \author{J{\O}rgen Christensen--Dalsgaard}
  \institute{Teoretisk Astrofysik Center, Danmarks Grundforskningsfond, and \\
  Institut for Fysik og Astronomi, Aarhus Universitet, \\
             DK--8000~Aarhus~C, Denmark }
\end{opening}

\begin{document}
\begin{abstract}
The overall framework for the study of solar convection and oscillations
is the spherically symmetric component of solar structure.
I discuss those properties of the solar interior
which depend on convection and other possible hydrodynamical motion
and the
increasingly detailed information about the structure which is provided
by helioseismic data.
The most basic dependence of solar models on convection is
the calibration to fix the solar radius.
The dominant causes for differences in oscillation frequencies
between the Sun and solar
models seem to be located near the top of the convection zone.
However, there is also
evidence for possible weak mixing below the convection zone and
perhaps in the solar core.
The former, at least, might be induced by penetration of convective
motion into the stable layers below.

\end{abstract}

\section{Introduction}
The outer solar convection zone extends over 29 \% of the solar
radius and contains about 2 \% of the mass of the Sun 
({\eg} Christensen-Dalsgaard, Gough \& Thompson 1991;
Kosovichev \& Fedorova 1991).
Within most of this region, energy transport is dominated
by convection, leading to a temperature gradient which
only deviates slightly from being adiabatic.
In particular, the structure of the convection zone is essentially
independent of the local value of the opacity.
Furthermore, matter is mixed on a time scale of months and hence
the composition may be assumed to be uniform.
In earlier phases of solar evolution, the convection zone has
extended considerably more deeply: thus it is normally assumed
that the Sun was fully convective before arriving on the main sequence,
justifying the assumption that the early Sun was chemically homogeneous.

Motion induced by convection is likely to extend beyond the
boundaries of the convection zone.
This can be observed in the solar atmosphere
and has a significant effect on the atmospheric structure.
Penetration beneath the lower boundary of the convection zone 
can only be inferred indirectly, but is potentially far more important for
overall solar structure and evolution.
It may affect the temperature stratification in the upper
parts of the radiative interior and cause mixing and transport
of angular momentum, either through direct motion in the form
of penetrating convective plumes or through convectively
induced gravity waves
({\eg} Schatzman, these proceedings; Zahn, these proceedings).
Clear evidence for such mixing is provided by the solar surface
abundances of lithium and beryllium which are considerably
reduced (by factors of about 140 and 2, respectively;
Anders \& Grevesse 1989),
relative to the initial composition of the solar system;
this indicates that matter has been mixed to a temperature
considerably higher than the maximum
temperature at the base of the convectively unstable region during the
main-sequence life of the Sun.

\begin{figure}[htbp]
\centerline{\psfig{figure=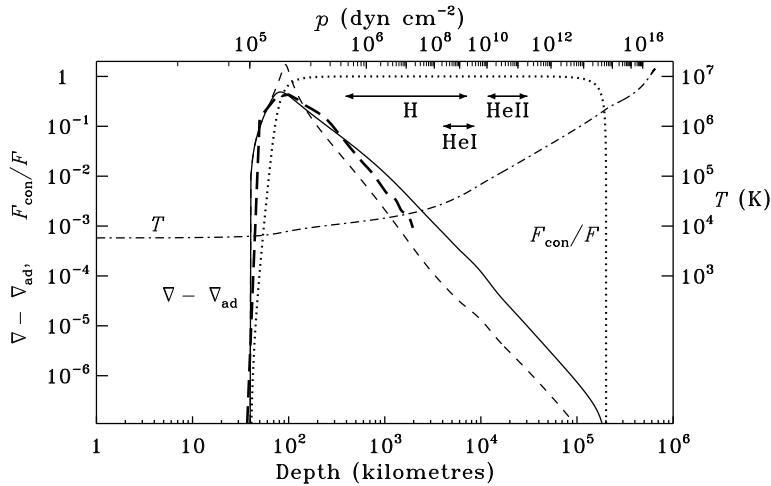,width=11cm}}
\caption{\label{gwcon} 
Properties of the solar convection zone.
The lower abscissa is depth below the point in the atmosphere
where the temperature equals the effective temperature,
whereas the upper abscissa is pressure $p$.
Most of the results are based on Model~S of 
Christensen-Dalsgaard {\etal} (1996),
which used the B\"ohm-Vitense (1958) mixing-length theory.
The dot-dashed curve, using the right-hand ordinate scale,
shows temperature $T$.
The remaining quantities refer to the left-hand ordinate scale.
The solid curve shows the superadiabatic temperature gradient
$\nabla - \nablaad$, 
where $\nabla = \dd \ln T/\dd \ln p$ and
$\nablaad = (\partial \ln T / \partial \ln p)_s$,
the derivative being at constant specific entropy $s$.
The dotted curve shows the convective flux $\Fcon$, in units
of the total flux $F$.
The horizontal arrows indicate the locations of the ionization
zone of hydrogen and the first and second ionization zones of helium,
extending between the points of 10 and 90 \% ionization.
In addition, the short-dashed curve shows $\nabla - \nablaad$
for a model using the Canuto \& Mazzitelli (1991) treatment of convection,
and the heavy long-dashed curve shows $\nabla - \nablaad$
in the average model resulting from hydrodynamical simulations
(Stein \& Nordlund 1989; Nordlund, these proceedings).
(Adapted from Gough \& Weiss 1976).
}
\end{figure}

Here I am concerned with the effects of convection 
on the spherically symmetric stellar structure and evolution,
and how these effects can be investigated through observations
of solar oscillations.
Many of these issues will be addressed in more detail in later
papers in the present volume;
however, I hope to provide a general framework,
as well as a basic impression of the data now available for
testing the solar models.

Gough \& Weiss (1976) pointed out that the properties of the
convection zone is essentially controlled by the thin,
substantially superadiabatic region at its top.
The integral of the superadiabatic gradient over this region
determines the adiabat of the nearly adiabatic part of the 
convection zone and hence its overall structure.
Provided the treatment of the superadiabatic region is adjusted,
{\eg} by varying suitable parameters, such as to yield the same
adiabat, the overall structure is insensitive to the details of that treatment.
Here I consider two simple parametrized treatments of convection.
One is the mixing-length theory of B\"ohm-Vitense (1958; in the following MLT),
with a mixing length proportional to the pressure scale height.
The second is the formulation by Canuto \& Mazzitelli (1991; CM),
with a characteristic scale related to the distance to the top of
the convection zone.
A potentially more realistic description of the superadiabatic region
can in principle be based on appropriate averages of
numerical solutions of the time-dependent 
hydrodynamical equations of convection.
I shall consider results of simulations carried out by Nordlund, Stein and
Trampedach ({\eg} Stein \& Nordlund 1989;
Nordlund, these proceedings; Trampedach {\etal}, these proceedings).
Unlike the simple formulations, this does not contain 
explicitly adjustable parameters;
hence it provides a prediction of the adiabat.

An overview of the structure of the solar convection zone
is provided by Fig.~\ref{gwcon}, in a form originally
introduced by Gough \& Weiss (1976).
This is based mostly on Model~S of Christensen-Dalsgaard {\etal} (1996);
the model was computed with the OPAL opacity 
(Iglesias, Rogers \& Wilson 1992)
and equation of state (Rogers, Swenson \& Iglesias 1996)
and included settling
and diffusion of helium and heavy elements,
using coefficients from Michaud \& Proffitt (1993).
Convection was treated using the MLT.
In addition, the figure shows superadiabatic gradients
obtained with the calibrated CM formulation and the 
hydrodynamical simulations.
It is evident that in all cases the region
of substantially superadiabatic convection is restricted to 
the outer few hundred kilometres of the convection zone.

\section{The structure of the solar convection zone}

As illustrated by Fig.~\ref{gwcon}
the region of significant superadiabaticity is extremely thin,
compared with the extent of the solar convection zone.
Thus the detailed structure of this region matters little
insofar as the overall structure of the star is concerned.
However, it provides the transition between the stellar
atmosphere and the almost adiabatic bulk of the convection zone.
The structure of the atmosphere can be found observationally,
in terms of semi-empirical atmospheric models.
Thus the integral over the superadiabatic gradient, which
determines the change in specific entropy between the atmosphere
and the interior of the convection zone, essentially fixes the
adiabat of the adiabatic part of the convection zone.
This, together with the equation of state and composition,
largely determines the structure of the convection zone.

The structure of the upper parts of the convection zone
is also affected by the dynamical effects of convection,
generally represented as a {\it turbulent pressure}
(see Rosenthal, these proceedings; Antia \& Basu, these proceedings).
These effects are often neglected in calculations of stellar models,
however.

\subsection{Properties of the convection zone}

To illustrate the properties of the convection zone 
it is instructive to consider a highly simplified model.
I assume the equation of state for a fully ionized perfect gas;
then the adiabatic relation between pressure $p$ and density $\rho$
can be written as
\be
p = K \rho^\gamma \; ,
\label{prho}
\ee
where $\gamma$ and $K$ may be assumed to be constant.
Neglecting also the mass contained in the convection zone
the equation of hydrostatic support is
\be
{\dd p \over \dd r} = - {G M \rho \over r^2} \; ,
\label{hydro}
\ee
where $r$ is distance to the centre of the star, $M$ is the mass
of the star and $G$ is the gravitational constant.
From equations~(\ref{prho}) and (\ref{hydro}) one obtains
\be
G M \left( {1 \over r} - {1 \over R_*} \right)
= K^{1/\gamma} {\gamma \over \gamma - 1} \left[p^{1 - 1/\gamma} -
p_*^{1 - 1/\gamma} \right] \; ,
\label{rsol}
\ee
where $p_*$ is the pressure at a point near the top of
the convection zone and $R_* \simeq R$ is the radius at this point,
$R$ being the surface radius of the star.

Conditions at the base of the convection zone are determined
by the transition to convective stability,
where matching to the radiative interior fixes the
radius $\rcz$ and pressure $\pcz$ at the convection-zone base.
The condition of marginal convective instability is
\be
{3 \over 16 \pi a \tilde c G} {\kappa p \over T^4} {L \over M} 
= \nablaad \; ,
\label{constab}
\ee
where $a$ is the radiation density constant, $\tilde c$ is the
speed of light, $T$ is temperature, $L$ is luminosity,
$\kappa$ is opacity, and we neglected again the mass in the convection zone;
also $\nablaad \simeq 2/5$.
This condition, together with equation~(\ref{prho}), the equation
of state and the dependence of $\kappa$ on $p$ and $T$,
determines the relation between $K$ and $\pcz$.
It is most simply analyzed by considering the response
of the model to a change $\delta K$ in $K$, keeping the other
parameters of the model, including mass and composition, fixed.
As confirmed by numerical computations, changes in the convective envelope
and outer part of the radiative interior have little effect on the
energy-generating core; thus $L$ is largely unchanged and so therefore,
according to equation~(\ref{constab}), is $\kappa p / T^4$.
Using the ideal gas law and equation~(\ref{prho}) we therefore obtain
\be
0 \simeq \delta \ln \left({\kappa p \over T^4} \right)
= {\kappa_T - 4 \over \gamma} \, \delta \ln K 
+ {1 \over \gamma} \left [ (\kappa_T - 4) ( \gamma - 1)
+ \gamma (\kappa_p + 1) \right] \, \delta \ln \pcz \; ,
\ee
where $\kappa_p = (\partial \ln \kappa / \partial \ln p)_T$
and $\kappa_T = (\partial \ln \kappa / \partial \ln T)_p$;
thus
\be
\delta \ln \pcz \simeq - {4 - \kappa_T \over (4 - \kappa_T) (\gamma - 1)
- \gamma(\kappa_p + 1)} \, \delta \ln K \; .
\label{delpcz}
\ee
At the base of the solar convection zone,
$\kappa_p \simeq 0.58$ and $\kappa_T \simeq -3.61$.
Thus, using $\gamma = 5/3$,
we find that $\delta \ln \pcz \simeq -3.1 \, \delta \ln K$.

We may now use equation~(\ref{rsol})
to find the resulting change in $R$.
Assuming that $\pcz \gg p_*$ and $R_* \simeq R$, we have that
\be
G M \left( {1 \over \rcz} - {1 \over R} \right)
\simeq K^{1/\gamma} {\gamma \over \gamma - 1} \pcz^{1 - 1/\gamma} \; .
\label{rcz}
\ee
The change in $\pcz$ evidently causes a change in $\rcz$;
assuming that the hydrostatic structure of the interior,
defined by $p(r)$, changes little up to the base of the convection zone,
\be
\delta \rcz \simeq - H_p \, \delta \ln \pcz \; ,
\label{delrcz}
\ee
where $H_p$ is the pressure scale height evaluated at the base of
the convection zone.
From equation (\ref{rcz}) it therefore follows that
\be
\delta R = - \left({R \over \rcz}\right)^2 H_p \, \delta \ln \pcz
+ {R \over \rcz} \dcz 
\delta \ln \left( K^{1/\gamma} \pcz^{1 - 1/\gamma} \right) \; ,
\label{delrs}
\ee
where $\dcz = R - \rcz$ is the depth of the convection zone.
Using again solar values, {\ie}, $\rcz \simeq 0.71 R$,
$H_p \simeq 0.081 R$, and the relation obtained above between
$\delta \ln \pcz$ and $\delta \ln K$, 
we find, separating the contributions from the two terms in
equation~(\ref{delrs})
\be
{\delta R \over R} \simeq 0.50 \, \delta \ln K - 0.26 \, \delta \ln K 
= 0.24 \, \delta \ln K \; ,
\label{delrsnum}
\ee
and hence
\be
{\delta \dcz \over R} \simeq -0.02 \, \delta \ln K \; .
\label{deldcznum}
\ee
It is remarkable, and perhaps surprising, that in the solar case
the depth of the convection zone appears to be virtually insensitive
to changes in the adiabat of the convection zone, the change
in surface radius resulting from the change
in the radius at the base of the convection zone.
As discussed below, this is confirmed by numerical results for solar models.
Note also that $H_p$ is approximately proportional to depth;
thus it is likely that the relative magnitude of the two terms
in equation~(\ref{delrs}), and hence the sign of the relation
in equation~(\ref{delrsnum}) will be roughly the same for other stars,
at least as long as the opacity derivatives do not change substantially.

\subsection{Calibration of the surface radius}

The dependence of $R$ on $K$ is used to calibrate solar models to have the
observed radius, by adjusting $K$.
This might most simply be achieved by assuming a discontinuity
in $\rho$ and $T$ at the top of the convection
zone such that $K$ attains the correct value ({\eg} Schwarzschild 1958).
However, more commonly a simplified physical description of
convection is used, generally containing a parameter
which can be adjusted to ensure the correct radius;
this might allow a safer extrapolation from the solar
case, where such calibration can be made, to models
of other stars where this is rarely possible.

The main features of this calibration can be illustrated by noting
that in the deeper part of the convection zone where matter is
essentially fully ionized, $K$ is related to the specific
entropy by
\be
s \simeq {3 \over 5} c_p \ln K \; ,
\label{entropy}
\ee
choosing the zero-point of entropy appropriately,
where $c_p$ is the specific heat at constant pressure,
and $\gamma$ was set to $5/3$.
The value $\sad$ of $s$ in adiabatic part of the convection zone
is related to the photospheric value $\sph$ by $\sad = \sph + \Delta s$,
where
\be
\Delta s = \int_{\ln \pph}^{\ln p^*} c_p (\nabla - \nablaad) \dd \ln p \; ;
\label{deltas}
\ee
here $p^*$ is a suitable point in the convection zone, such that
$\nabla - \nablaad \ll 1$.
Assuming that the atmospheric structure is approximately
unchanged, the change in $K$ is obtained from
the change in $\Delta s$ as
\be
\delta \ln K \simeq {5 \over 3 } {\delta (\Delta s) \over c_p} \; .
\label{delK}
\ee
For a given energy flux, $\nabla - \nablaad$
is determined by the efficacy of convection in the superadiabatic
region: if convective transport becomes more efficient,
the superadiabatic gradient is reduced, and so therefore
is $\Delta s$ and hence $K$,
which according to equation~(\ref{delrsnum})
leads to a smaller radius of the model.

Several formulations of convection, including the commonly used
mixing-length theory, determine the convective efficacy
in terms of a characteristic scale $\ell$,
such as the size or mean free path of a convective element.
This is often parametrized as a multiple $\alphac$ of 
a typical length scale in the model, such as the local
pressure scale height or the distance to the boundary of
the convection zone.
In the limit of efficient convection, relevant to the larger part
of the solar convection zone, the convective flux then satisfies
$\Fcon \propto (\nabla - \nablaad)^{3/2} \alphac^2$.
It follows that if the luminosity, and hence approximately
the convective flux, is kept fixed,
$\nabla - \nablaad \propto \alphac^{-4/3}$.
According to equations~(\ref{deltas}) and (\ref{delK})
we therefore have that
\be
\delta \ln K \simeq -2 {\Delta s \over c_p} \delta \ln \alphac  
\simeq - \delta \ln \alphac \; ,
\ee
where the last equality used solar values for $\Delta s$;
hence, from equation~(\ref{delrsnum}),
\be
\delta \ln R \simeq -0.24 \delta \ln \alphac \; .
\label{delrsal}
\ee

\subsection{Results for stellar models}

To illustrate the behaviour of convective envelopes 
discussed in sections 2.1 and 2.2,
I have calculated three static models, based on the composition
profile of Model~S (Christensen-Dalsgaard {\etal} 1996).
Some properties of the models are summarized in Table~1.
Models~1 and 2 have been calibrated to solar radius and luminosity
by adjusting the convective efficacy and scaling the hydrogen
abundance, as a function of mass, by a suitable factor 
({\eg} Christensen-Dalsgaard \& Thompson 1991).
In all cases, the envelope hydrogen abundance by mass
is $X_{\rm e} \simeq 0.737$.
Model~1 used a version of the Canuto \& Mazzitelli (1991)
convection formulation, with mixing-length $\ell$ proportional
to the distance to the convection-zone boundary, but including
a parameter which allows calibration to the precise solar radius.
Model~2 used the B\"ohm-Vitense mixing-length treatment,
with a mixing length $\ell = \alphac H_p$
proportional to the pressure scale height.
Model~3 is also based on the B\"ohm-Vitense formulation,
but choosing a different mixing length; the composition is the same
as for Model~2.
For these models turbulent pressure was ignored.
In addition, the table lists an envelope model (Model~4)
obtained by matching to a hydrodynamical simulation,
in the manner of Trampedach {\etal} (these proceedings);
convection was treated using MLT, with $\alphac$ and a factor in
the treatment of turbulent pressure adjusted to obtain a continuous
match of pressure and density in the deepest part of the simulation.
Thus the adiabat of the deep convection zone is fixed by the
properties of the simulation.
For this model the hydrogen abundance is $X_{\rm e} \simeq 0.703$.

\begin{table}[htbp]
  \caption[]{\label{tab_1}     
   Properties of static solar models treating
   convection with the Canuto \& Mazzitelli
   formulation (CM) or the B\"ohm-Vitense mixing-length
   formulation (MLT), as well as an envelope matched continuously
   to results of hydrodynamical simulation (SIM).
   The radius $R$ and $\dcz$ of the convection
   zone are given in units of the solar radius 
   $R_\odot = 6.9599 \times 10^{10} \cm$.
   For the remaining notation, see text.
        }     
  \begin{center}
   \begin{tabular}{cccccccc}  %
     \hline
     Model &  & $\alphac$ & $R / R_\odot$ & $\dcz / R_\odot$ & 
     $\Delta s$ & $K$ & $\pcz$ \\  
      &  &  &  &  & (cgs) & (cgs) & $\dyn \cm^{-2}$ \\
     \hline    
     1 & CM & -- & 1.0000 & 0.2886 & $1.881 \times 10^8$ &
     $9.12 \times 10^{14}$ & $5.76 \times 10^{13}$ \\   
     2 &MLT & 1.9894 & 1.0000 & 0.2884 & $1.881 \times 10^8$ &
     $9.12 \times 10^{14}$ & $5.74 \times 10^{13}$ \\   
     3 &MLT & 1.7894 & 1.0226 & 0.2883 & $2.069 \times 10^8$ &
     $9.87 \times 10^{14}$ & $4.46 \times 10^{13}$ \\   
     4 &SIM & -- & 1.0000 & 0.3037 & $1.953 \times 10^8$ &
     $7.49 \times 10^{14}$ & $9.19 \times 10^{13}$ \\   
     \hline
   \end{tabular}
  \end{center}
\end{table}

\begin{figure}[htbp]
\vskip -0.5truecm
\centerline{\psfig{figure=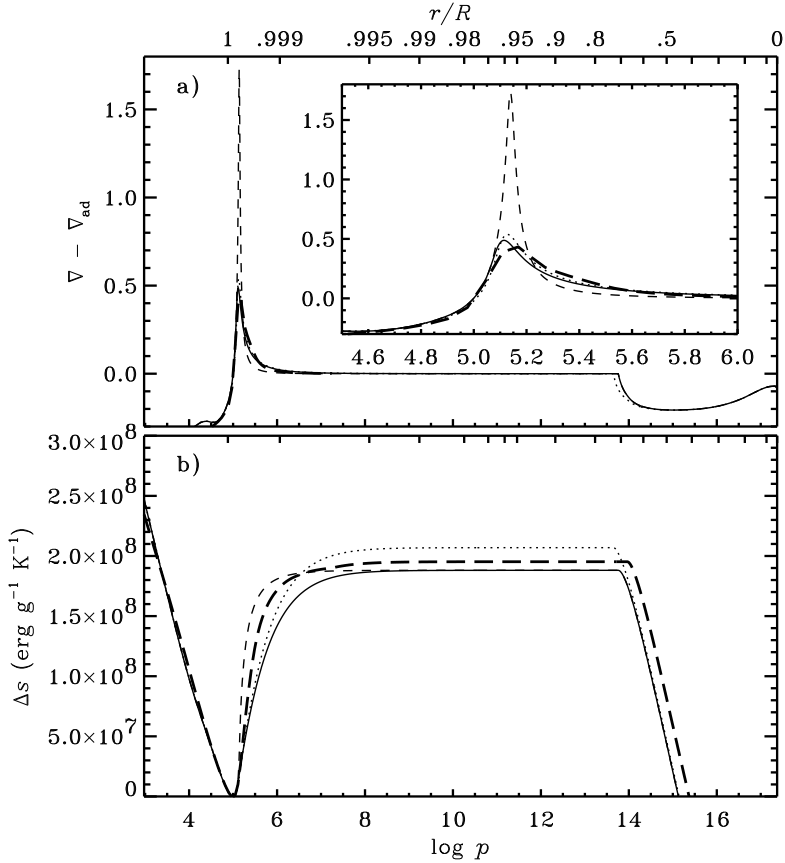,width=10.5cm}}
\vskip -0.5truecm
\caption{\label{superad} 
Superadiabatic gradient (a) and entropy difference $\Delta s$ (b) in
the four models listed in Table~1,
using the following line styles:
Model~1 (calibrated CM formulation) short-dashed line;
Model~2 (calibrated MLT formulation) solid line;
Model~3 (uncalibrated MLT formulation) dotted line;
Model~4 (hydrodynamical simulation matched to MLT envelope) heavy 
long-dashed line.
The lower abscissa is logarithmic (to base 10) of pressure,
whereas the upper abscissa shows the corresponding
fractional radius in Model~2.
The insert in panel (a) shows details of the superadiabatic peak.
In panel (b) the zero-point of $\Delta s$ has been chosen at the top
of the convection zone.
}
\end{figure}

Figure~\ref{superad} shows the superadiabatic gradient and
the integrated entropy change, calculated from equation~(\ref{deltas}).
It is evident that the CM formulation leads to a much higher and
sharper peak in $\nabla - \nablaad$, with a corresponding 
strongly confined change in $\Delta s$.
However, with the calibration to solar radius, the
value of the entropy in the adiabatic part of the convection zone,
and hence the properties of the interior of the model, are
essentially the same in Models~1 and 2, as shown in more detail
by the results presented in Table~1.
On the other hand, the decrease in $\alphac$ in Model~3, relative
to Model~2, causes an increase in $\nabla - \nablaad$,
a corresponding increase in $\Delta s$ and hence an increase in the
radius of the model.
The magnitude of the relative change in $R$, 2.3~\%, is quite
close to what is predicted by the simple approximation~(\ref{delrsal}).
Also, it should be noticed that the depth $\dcz$ of
the convection zone has changed little, in accordance
with equation~(\ref{deldcznum}).

The results for the hydrodynamical simulation, matched to an envelope
model, cannot be interpreted as simply in terms of the results
of sections~2.2 and 2.3.
Equation~(\ref{rcz}) still holds; however, since for the envelope
model $R$ is fixed, the relation defines the change in the
depth of the convection zone, as shown in Table~1.
In this case one cannot assume that the interior is unaffected
by the change in $K$, as was the case for complete models,
and hence equation~(\ref{delrcz}) is no longer valid.
Also, because of the difference in composition between Models~2 and 4,
say, there is no longer a simple connection between the changes
in $\Delta s$ and $K$.
These questions deserve more careful study than can be attempted here,
not least in connection with the calibration,
described by Trampedach {\etal} (these proceedings),
of the mixing length on the basis of convection simulations.
It should be noticed also that the matched envelope predicts
a convection zone extending somewhat more deeply than in the
calibrated models, as a result of the different value of $K$.
Nevertheless, given the fact that no attempt was made in the simulation
to match the solar adiabat, 
it is encouraging that the change in $K$ is relatively modest.
Comparisons such as the one attempted here are undoubtedly
important tests of simulations of solar convection. 

\subsection{How adiabatic is the deep part of the convection zone?}

The approximation used in equation~(\ref{prho}) is only valid
if $\gamma$ is constant.
However, to the extent that the stratification can be assumed
to be adiabatic, $p$ and $\rho$ are related by
\be
{\dd \ln p \over \dd \ln \rho} = \gamma_1 
\equiv \left( {\partial \ln p \over \partial \ln \rho} \right)_s \; ,
\label{prhoad}
\ee
where the dependence of $\gamma_1$ on $p$, $\rho$ and composition
is determined by the equation of state of the gas.
It follows that the structure of the adiabatic part of the
convection zone is entirely specified by the equation of state,
the composition and the actual value $\sad$ of the specific entropy.
This property makes the adiabatic part of the convection zone
a valuable tool for investigations of the
equation of state of stellar material
({\eg} Christensen-Dalsgaard \& D\"appen 1992)
and forms the basis for helioseismic determinations of the solar
envelope helium abundance ({\eg} Kosovichev {\etal} 1992).

It is evident that these analyses are possible only to
within the accuracy of equation~(\ref{prhoad}).
Since the thermodynamic effects under consideration are
minute, this imposes severe constraints on the
superadiabatic gradient.
The behaviour for three different treatments of convection,
in relation to the location of the dominant ionization zones,
was illustrated in Fig.~\ref{gwcon}.
In the hydrogen and, at least for MLT,
the first helium ionization zone $\nabla - \nablaad \sim 10^{-3}$,
which is comparable with the effects introduced by current uncertainties
in the equation of state.
Thus investigations of the equation of state have generally been 
concentrated on the second helium ionization zone, where
$\nabla - \nablaad$ is at most around $10^{-4}$ in the MLT model.
According to the CM formulation,
where convective efficiency
increases rapidly with depth, the allowable range might
include also the first helium ionization zone.
The hydrodynamical simulations do not extend sufficiently deeply
to reach the second helium ionization zone but appear, from
Fig.~\ref{gwcon}, to yield a superadiabatic gradient intermediate
between MLT and CM.
These substantial differences 
indicate that the uncertainty in the treatment of convection
might have a significant influence on the degree of adiabaticity
even in the second helium ionization zone.
These issues need further investigation, before very detailed
tests of the equation of state and/or precise determination of
the helium abundance can be carried out.

I finally note that turbulent pressure might also
influence tests of the thermodynamic properties of the solar plasma.
For example, the ratio between the turbulent and total pressure 
at the upper edge of the second helium ionization zone in the MLT Model~2
can be estimated as approximately $10^{-4}$.
This could have significant effects on the relation between 
$\gamma_1$, $p$, $\rho$ and composition inferred from helioseismic analyses.

\section{Helioseismic data on solar models}

Extensive reviews on solar oscillations and their application
to helioseismology were given, {\eg}, by 
Gough \& Toomre (1991), Christensen-Dalsgaard \& Berthomieu (1991),
Gough \& Thompson (1991) and Gough (1993).
Since I consider only the spherically symmetric structure,
the oscillation frequencies $\omega_{nl}$ depend on the degree $l$
and radial order $n$ alone.
I shall assume the adiabatic approximation, neglecting
the energy gain or loss of the oscillations.
This is an excellent approximation in almost the entire Sun
but breaks down in the near-surface region where the thermal
time scale becomes comparable with the oscillation period.
As discussed in detail by Rosenthal (these proceedings)
this region gives rise to other uncertainties in the
computation of the frequencies,
arising from the physics of the model and the oscillations;
thus in any case the presence of errors in the computed frequencies,
arising from the near-surface region, must be kept in mind.

The observed frequencies correspond mostly to acoustic modes.
These are trapped between an upper turning point just below
the photosphere and a lower turning point, at a distance $\rt$
from the centre determined by
\be
{c(\rt) \over \rt} = {\omega \over l + 1/2} \; ,
\label{lowturn}
\ee
where $c$ is the adiabatic sound speed.
It follows from equation~(\ref{lowturn}) that high-degree
modes are trapped near the solar surface whereas low-degree modes
penetrate into the solar core.

\begin{figure}[htbp]
\vskip -0.3cm
\centerline{\psfig{figure=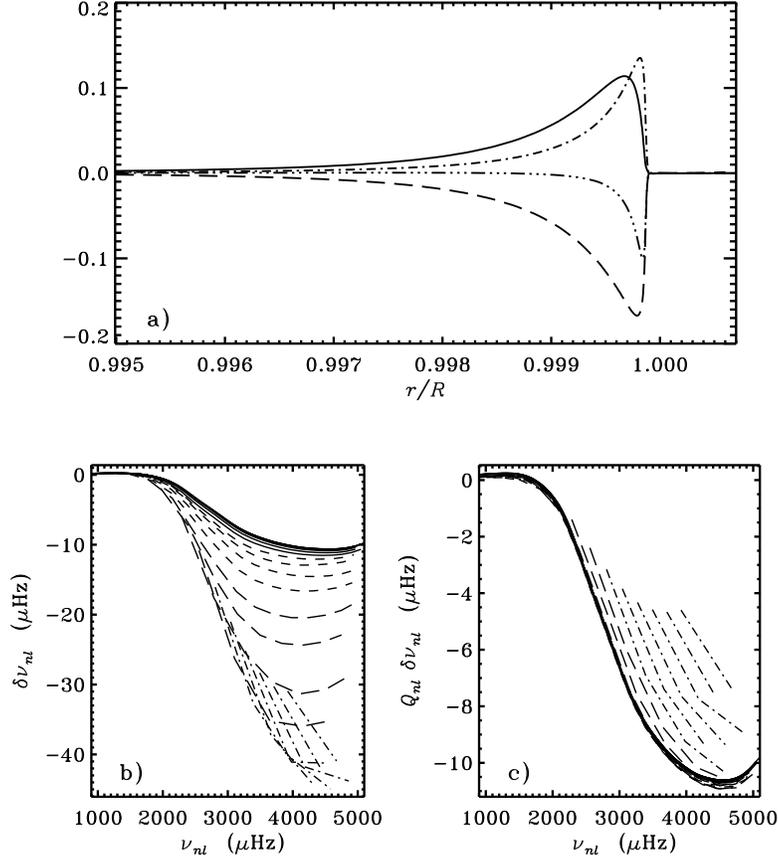,width=10.5cm}}
\caption{\label{condiff} 
(a) Differences, at fixed mass, between Model~1 and Model~2 
of Table~1, computed with
the CM and the MLT convection formulations, in the sense (CM) -- (MLT).
The following differences are shown:
$\delta \ln c^2$ (solid line);
$\delta \ln \rho$ (dashed line);
$\delta \ln T$ (dot-dashed line); and
$\delta \ln \gamma_1$ (triple-dot-dashed line).
(b) Unscaled frequency differences between the same two models,
for the following degrees:
$l = 0 - 30$ (solid lines);
$l = 40, 50, 70, 100$ (short dashed lines);
$l = 150, 200, 300, 400$ (long dashed lines); and
$l =  500, 600, 700, 800, 900, 1000$ (dot-dashed lines).
(c) Corresponding scaled frequency differences, using the same line styles.
}
\end{figure}

Perturbation analysis of the oscillation equations
({\eg} Christensen-Dals\-gaard \& Berthomieu 1991; see also 
Rosenthal, these proceedings) shows that the near-surface
effects cause changes in the frequencies of the form
\be
\delta \omega_{nl}^{\rm (ns)} = {\Fsurf (\omega_{nl}) \over Q_{nl}} \; ,
\label{delomns}
\ee
for modes of low or moderate degree.
Here $Q_{nl} = E_{nl} / \bar E_0(\omega_{nl})$,
where $E_{nl}$ is the mode energy, normalized by the squared
surface amplitude, and $\bar E_0 (\omega_{nl})$ is the energy
of a radial ($l = 0$) mode, interpolated to $\omega_{nl}$.
Thus the behaviour of $Q_{nl}$ reflects the variation of
the turning-point radius $\rt$ with the degree of the mode:
high-degree modes involve a smaller part of the Sun than do
low-degree modes and therefore have smaller normalized energy and 
$Q_{nl}$, hence according to equation~(\ref{delomns})
making their frequencies more susceptible to the near-surface effects.
$\Fsurf(\omega)$ is a function of frequency which depends on
the physics of the near-surface region; it may be shown
that if the errors in the calculation are confined extremely
close to the surface, $\Fsurf$ is a slowly varying function
of $\omega$ which is small at low frequency.

Equation~(\ref{delomns}) motivates 
analyzing frequency differences in terms of 
$Q_{nl} \delta \omega_{nl}$.
This scaling effectively reduces the frequency change 
resulting from near-surface effects to the equivalent change for 
a radial mode, by taking out the dependence on the penetration depth.
Thus if differences in structure were confined
exclusively to the near-surface region, we might expect
$Q_{nl} \delta \omega_{nl} $ to depend on frequency alone,
for modes of low or moderate $l$ for which the motion in the
surface layers is almost radial.

These principles may be illustrated 
by comparing Models~1 and 2 computed with the CM and
MLT treatments of convection and, according to
Fig.~\ref{superad}, differing only very near the surface.
Figure~\ref{condiff}a shows
differences, at fixed mass, between these models;
as discussed by Christensen-Dalsgaard \& Thompson (1996)
the effects of near-surface model changes are most naturally
represented in terms of such differences.
It is evident that the change in the model is essentially
confined to the superadiabatic region where the temperature
gradients differ substantially.
Unscaled and scaled frequency differences between these
two models are shown in panels (b) and (c) of Fig.~\ref{condiff}.
The unscaled differences show a fairly substantial dependence
on degree which is largely suppressed by the scaling,
except at high degree.

\begin{figure}[htbp]
\vskip -0.7cm
\centerline{\psfig{figure=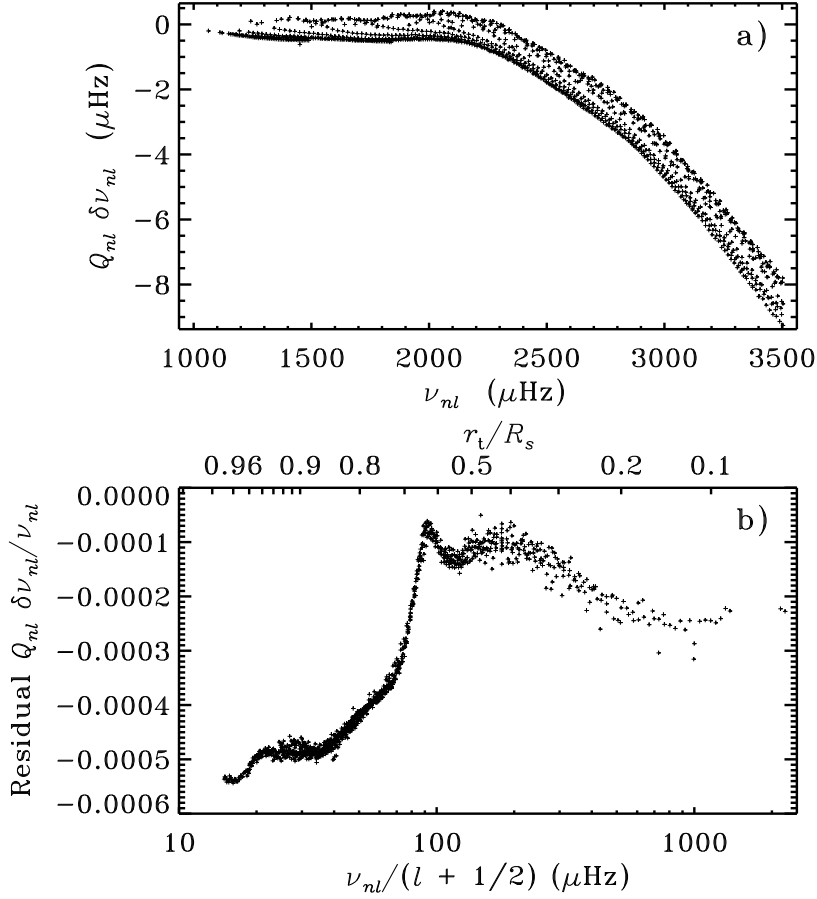,width=11cm}}
\caption{\label{obsdiff} 
(a) Scaled differences between cyclic frequencies $\nu = \omega/2 \pi$
observed with
the LOWL instrument (Tomczyk {\etal} 1995) and frequencies of
Model~S of Christensen-Dalsgaard {\etal} (1996), in the sense
(Sun) -- (Model).
Modes of degree $l = 0 - 99$ are included.
(b) Residual scaled relative frequency differences, after subtraction
of the function $\CHtwo(\omega)$ obtained from a fit of the
form~(\ref{diffduv}).The upper abscissa shows the turning-point radius
$\rt$, related to $\omega/(l + 1/2)$ through equation~(\ref{lowturn}).
}
\end{figure}

It is convenient to relate frequency differences between two models,
or between the Sun and a model, to the corresponding differences in structure.
This is generally done on the assumption 
that the differences are sufficiently small that the relation is linear.
Asymptotic theory for acoustic modes then shows that
(Christensen-Dalsgaard, Gough \& P\'erez Hern\'andez 1988)
\be
S_{nl} {\delta \omega_{nl} \over \omega_{nl}}
= \CHone \left( {\omega_{nl} \over l+1/2} \right)
+ \CHtwo(\omega_{nl}) \; ,
\label{diffduv}
\ee
where $S_{nl}$ is closely related to the scaling $Q_{nl}$ 
introduced in equation~(\ref{delomns}),
\be
\CHone(w) = \int_{\rt}^R \left( 1-{c^2\over{r^2 w^2}} \right)^{-1/2}
{\delta_r c\over c}{\dd r\over c} \; , 
\label{diffduv-h}
\ee
and the function $\CHtwo(\omega)$ contains contributions from the 
near-surface region, including the uncertain aspects of the physics.
In equation~(\ref{diffduv-h}) the difference $\delta_r c$ 
is evaluated at fixed radius $r$.

Figure~\ref{obsdiff}a shows differences between the first year's
observations with the LOWL instrument
(Tomczyk {\etal} 1995; Tomczyk, Schou \& Thompson 1996)
and Model~S of Christensen-Dalsgaard {\etal} (1996),
scaled as suggested by equation~(\ref{delomns}).
Clearly the frequency differences depend primarily on frequency.
This suggests that the differences between the Sun and the model
are dominated by the near-surface effects.
Indeed, the shape and magnitude of the differences are rather similar to
the differences, shown in Fig.~\ref{condiff}, between
the CM and MLT models.
This might suggest that the CM formulation may be
a more accurate representation of the uppermost layers of
the solar convection zone ({\eg} Patern\`o {\etal} 1993);
however, as discussed by Rosenthal (these proceedings)
this conclusion may be premature, given the other potential
near-surface contributions to the frequency differences.

Equation~(\ref{diffduv}) indicates that the relative frequency differences
can be separated into a contribution depending on frequency and
a contribution depending on $\omega/(l + 1/2)$ or, equivalently,
the turning-point radius $\rt$.
Indeed, by fitting this relation to relative differences
corresponding to those shown in Fig.~\ref{obsdiff}a and
subtracting the component corresponding to $\CHtwo(\omega)$
one obtains the residuals shown in Fig.~\ref{obsdiff}b
which are clearly predominantly a function of $\omega/(l + 1/2)$.
The most striking feature is the rapid variation for modes
whose turning points are near $0.7 R$, {\ie}, close to the
base of the convection zone.
According to equation~(\ref{diffduv-h}) this suggests 
that there is a
comparatively large component of the sound-speed difference at this
point; as we shall see in the next section, this is in fact the case.

\section{Structure of the radiative interior}

To investigate the causes of the residual frequency
differences shown in Fig.~\ref{obsdiff}b a more careful analysis is required.
In particular, it is preferable to move
beyond the asymptotic approximation in equation~(\ref{diffduv}).
From general properties of the oscillation equations
one may express small differences in adiabatic frequency
linearly in terms of
differences in two variables characterizing the model,
{\eg} $\delta_r c^2$ and $\delta_r \rho$
({\eg} Gough \& Thompson 1991).
The actual differences between the observed frequencies
and adiabatic frequencies of a model must also reflect
the nonadiabatic effects on the frequencies and the inadequacies
in the modelling of the near-surface region.
As a result, the frequency differences can be expressed as 
\be
{\delta \omega_{nl} \over \omega_{nl}} =
\int_0^R \left[ K_{c^2, \rho}^{nl}(r) {\delta_r c^2 \over c^2}(r)
+ K_{\rho,c^2}^{nl}(r) {\delta_r \rho \over \rho}(r) \right] \dd  r
+ Q_{nl}^{-1} \CG(\omega_{nl}) + \epsilon_{nl} \; ,
\label{numdiff}
\ee
$\epsilon_{nl}$ being the observational error;
here the kernels $K_{c^2, \rho}^{nl}$ and $K_{\rho,c^2}^{nl}$
are determined from the
eigenfunctions in the model, while the penultimate term arises
from the neglected physics in the near-surface region.

\begin{figure}[htbp]
\vskip -0.3cm
\centerline{\psfig{figure=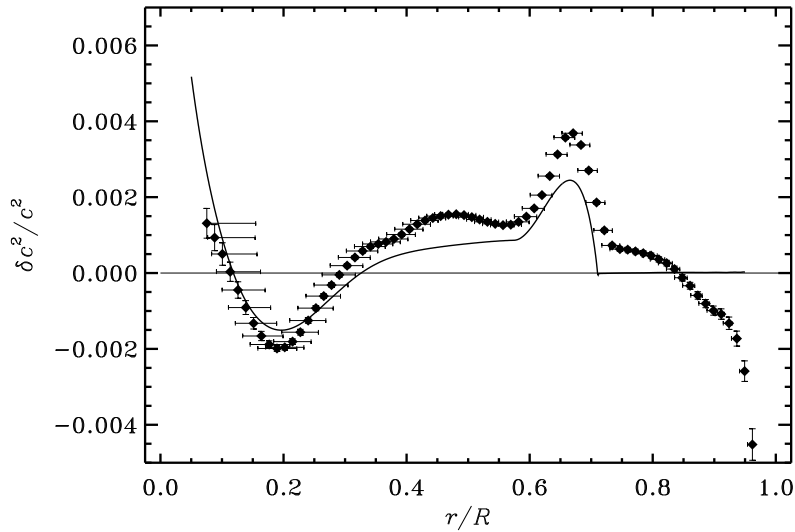,width=11cm}}
\caption{\label{fig:delcinv} 
The symbols show results of SOLA inversion 
(Basu {\etal} 1996b) of frequencies
obtained with the
the LOWL instrument (Tomczyk {\etal} 1995), to infer the
sound-speed difference between the Sun and
Model~S of Christensen-Dalsgaard {\etal} (1996), in the sense
(Sun) -- (Model).
The vertical error bars show the standard error in the result,
obtained from the estimated errors in the observed frequencies,
whereas the horizontal bars indicate the extent of the averaging
kernels ({\cf} equation~\ref{avker}).
The curve shows the change in squared sound speed resulting
from the modification to the hydrogen profile illustrated in Fig.~\ref{xprof}
({\cf} Bruntt 1996).
}
\end{figure}

Equation~(\ref{numdiff}) forms the basis for inverse analyses to
infer properties of $\delta_r c$ and $\delta_r \rho$.
Here I consider the so-called subtractive optimally localized 
averages (SOLA) technique, first introduced by Pijpers \& Thompson (1992);
details of the implementation were provided by 
Basu {\etal} (1996a).
The principle is to construct linear combinations of equation~(\ref{numdiff})
with weights $c_{nl}(r_0)$ chosen such that the {\it averaging kernel}
\be
\CK_{c^2, \rho}(r_0, r) =
\sum_{nl} c_{nl} (r_0) K_{c^2, \rho}^{nl}(r) 
\label{avker}
\ee
is a function localized near $r \simeq r_0$,
whereas the remaining terms on the right-hand side of equation~(\ref{numdiff})
are minimized.
In particular, the contribution
from the near-surface problems, as given by the term in $\CG(\omega)$,
can largely be eliminated.
To the extent that this procedure is successful, 
we obtain a localized estimate of $(\delta_r c/c)(r_0)$,
\be
\overline{ \left( {\delta_r c \over c} \right)} (r_0) =
\sum_{nl} c_{nl} (r_0) {\delta \omega_{nl} \over \omega_{nl}} \simeq
\int_0^R \CK_{c^2, \rho} (r_0, r) {\delta_r c^2 \over c^2}(r) \dd r \; .
\label{delcinv}
\ee
An estimate of the standard error in the result can be obtained
from the observational standard deviations of $\epsilon_{nl}$.

\begin{figure}[htbp]
\vskip -0.2cm
\centerline{\psfig{figure=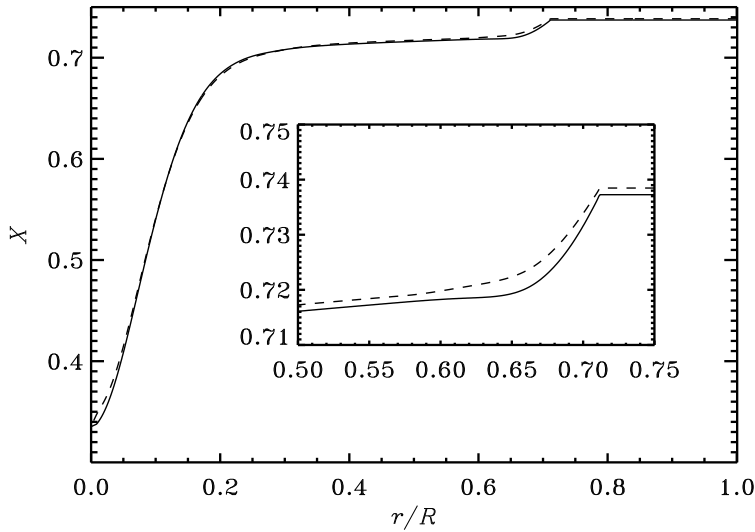,width=11cm}}
\caption{\label{xprof} 
Profiles of the abundance $X$ by mass of hydrogen.
The solid line shows the profile in
Model~S of Christensen-Dalsgaard {\etal} (1996),
whereas the dashed line shows a modified profile 
aimed at trying to 
match the sound-speed difference shown in Fig.~\ref{fig:delcinv}
between the Sun and the model.
The insert provides a blow-up of the region around the base of
the convection zone.
Adapted from Bruntt (1996).
}
\end{figure}

Figure~\ref{fig:delcinv} shows the difference in squared sound speed 
resulting from an application of this procedure to the 
frequency differences illustrated in Fig.~\ref{obsdiff}a
(Basu {\etal} 1996b).
It is evident that there is indeed a sharp feature in the
sound-speed difference just below the convection zone;
this is responsible for the behaviour of the residual
frequency differences around $\nu/(l+1/2) \simeq 80 \muHz$
in Fig.~\ref{obsdiff}b.
A second substantial feature is the dip in $\delta_r c^2/c^2$
around $r \simeq 0.2 R$, {\ie}, at the edge of the nuclear-burning
core, apparently followed by a rise in the deeper parts of the core.
In the convection zone the difference is relatively small, although
with a rise in magnitude towards the surface;
this might be caused by errors in the equation of state or possibly
by residual effects of the near-surface problems.

To discuss the possible causes for the sound-speed differences
in the radiative interior I note that, according to the ideal gas law,
\be
c^2 = {\gamma_1 p \over \rho} 
\simeq {\gamma_1 k_{\rm B} T \over \mu m_{\rm u}} \; ,
\label{capprox}
\ee
where $k_{\rm B}$ is Boltzmann's constant, 
$\mu$ is the mean molecular weight and 
$m_{\rm u}$ is the atomic mass unit.
Thus $\delta_r c/c$ must reflect a difference in $T / \mu$ between the Sun
and the model.
With this in mind, it is striking that the two regions of
dramatic variation in $\delta_r c/c$ coincide with
regions where the composition, and hence $\mu$, varies strongly.
This is illustrated by the hydrogen profile shown in Fig.~\ref{xprof}:
beneath the convection zone the accumulation of helium
settling out of the convection zone causes a sharp gradient in $X$,
whereas hydrogen burning, with an additional small contribution
from helium settling, leads to a strong variation of $X$ in the core.
In both cases, the difference between the solar and model sound speed
could be reduced by smoothing the composition profile:
this would increase $X$, reduce $\mu$ and hence increase $c$
just below the convection zone and similarly reduce $X$ and $c$
at the edge of the core, with a corresponding increase in the inner core.

To test this possibility, Bruntt (1996) adjusted the
hydrogen profile in Model~S in such a way as to approximate
the sound-speed difference shown in Fig.~\ref{fig:delcinv}.
The profiles were constrained to correspond to
the same total amount of hydrogen as for Model~S,
to within 0.5 \%, and to give the observed solar luminosity;
however, no assumptions were made about possible physical
mechanisms which might cause the redistribution of hydrogen.
The resulting profile is shown in Fig.~\ref{xprof} as a
dashed line, and the difference in sound speed between the
modified model and Model~S was shown 
as the solid line in Fig.~\ref{fig:delcinv}.
Evidently the change in composition has
reproduced much of the sound-speed difference between the Sun and the model.
Evidence for a smoother composition profile just below the convection zone
was also found from inverse analysis by Antia \& Chitre (1996).

Such changes in composition are not implausible.
Indeed, as mentioned in the introduction, the depletion of lithium
and beryllium demonstrates that mixing in the Sun well below
the convection zone must have taken place at some stage during
solar main-sequence evolution.
Computations with rotationally-induced mixing 
accounting for the lithium depletion (Chaboyer {\etal} 1995;
Richard {\etal} 1996; see also Zahn, these proceedings)
have largely succeeded in eliminating the bump in the sound-speed
difference just below the convection zone (Gough {\etal} 1996).
Mixing would also be caused by convective penetration into the
stable region; even a very small fraction of convective eddies
penetrating to a substantial distance could cause appreciable
mixing, with little effect on the temperature structure.
It has furthermore been suggested that gravity waves excited by 
convective penetration
might lead to mixing ({\eg} Montalb\'an 1994; Montalb\'an \& Schatzman 1996;
see also Schatzman, these proceedings).

It should be noted that the lithium depletion and
change in the hydrogen profile are not automatically related:
thus strong mixing in the early phases of solar evolution,
as might have been caused by rotation, could
have depleted lithium with little effect on the present hydrogen profile.
In this sense, the information obtained from lithium and from
the sound-speed inversion is complementary.

Unfortunately, mixing is not the only mechanism that might account
for the sound-speed results.
Early mass loss ({\eg} Guzik \& Cox 1995)
might also reduce the lithium and beryllium abundances,
and in addition improve the agreement with the inferred solar
sound speed just below the convection zone (Gough {\etal} 1996).
Furthermore, it is obviously possible to change the sound speed
by changing the temperature profile.
This requires modifications to the opacity such that the
condition of radiative energy transport is satisfied.
Tripathy, Basu \& Christensen-Dalsgaard (1996)
showed that the inferred sound-speed difference in
Fig.~\ref{fig:delcinv} can be reproduced by means of a suitably
chosen opacity modification of only a few per cent.
Thus independent information and, if possible, tighter
constraints on the opacity are required to separate the
different possible causes for the remaining differences
between the Sun and solar models.

\section{Discussion}

The results of inversion for the sound-speed difference,
such as those shown in Fig.~\ref{fig:delcinv},
indicate a strikingly
close agreement between the solar sound speed and that of normal solar models.
This has little implication for the dynamics of the upper parts
of the convection zone which have been adjusted, by calibration of
the mixing length, to produce a model with the correct radius;
however, it does indicate that conditions at and below
the base of the convection zone are not vastly different from
those obtained from normal stellar modelling
(see also Roxburgh, these proceedings).
Nonetheless, the most dramatic sound-speed difference does in fact
occur in this region.
Although various explanations are possible, the most plausible of these
is perhaps mixing induced by rotational instability,
direct convective penetration or gravity waves.
However, it is important to stress that, despite
its great power, helioseismology cannot on its own provide a
full investigation of the problems of mixing in stellar interiors.
This requires a combination of a better physical understanding
of the mixing processes, data from other stars on, for example,
the lithium depletion ({\eg} Michaud \& Charbonneau 1991),
and tighter constraints on other aspects of the physics of
the radiative interior, such as the opacity.

It is encouraging that the hydrodynamical simulations discussed
here result in a structure of the adiabatic part
of the convection zone fairly close to that obtained in
calibrated parametrized models.
This offers some hope that such simulations might be used to
provide a firmer extrapolation to other stars than the commonly used
assumption of a constant mixing-length parameter
(Ludwig, Freytag \& Steffen, these proceedings;
Trampedach {\etal}, these proceedings).
Tests of such extrapolations might be provided by
well-observed binary stars, such as the $\alpha$ Cen system
({\eg} Edmonds {\etal} 1992; Fernandes \& Neuforge 1995).
Observations of solarlike oscillations in these or other stars
might clearly be extremely valuable in constraining the models,
in terms of the properties of convection or other aspects of the structure.
When such data are available, the meaning of the ``S'' in the title
of subsequent conferences in this series might be subtly changed.

\footnotesize
\section*{\bf Acknowledgements}

I am grateful to Hans Bruntt for permission to show
the results of the modified hydrogen profile in Figs~\ref{fig:delcinv} and
\ref{xprof},
to Regner Trampedach for the matched envelope and hydrodynamical
simulation and to Mario Monteiro for the implementation of the
CM formalism.
Colin Rosenthal is thanked for comments on an earlier version of the paper.
This work was
supported by the Danish National Research Foundation through its establishment
of the Theoretical Astrophysics Center.


\section*{References}

\bibref
Anders E. \& Grevesse N., 1989.
{\it Geochim. Cosmochim. Acta}
{\bf 53}, 197 
\bibref
Antia H. M. \& Chitre S. M., 1996.
Submitted to {\it ApJ}.
\bibref
B\"ohm-Vitense E., 1958.
{\it Z. Astrophys.}
{\bf 46}, 108 
\bibref
Basu S., Christensen-Dalsgaard J.,
P\'erez Hern\'andez F. \& Thompson M. J., 1996a.\\
{\it MNRAS} {\bf 280}, 651 
\bibref
Basu S., Christensen-Dalsgaard J., Schou J., Thompson M. J. \&
Tomczyk S., 1996b.
{\it Bull. Astron. Soc. India} {\bf 24}, 147 
\bibref
Bruntt H., 1996.
{\it Batchelor thesis}, Aarhus University.
\bibref
Canuto V. M. \& Mazzitelli I., 1991.
{\it ApJ} {\bf 370}, 295 
\bibref
Chaboyer B., Demarque P., Guenther D. B. \& Pinsonneault M. H., 1995.
{\it ApJ} {\bf 446}, 435 
\bibref
Christensen-Dalsgaard J. \& D\"appen W., 1992.
{\it A\&AR\/} {\bf 4}, 267.
\bibref
Christensen-Dalsgaard J. \& Berthomieu G., 1991.
In {\it Solar interior and atmosphere}, 
eds Cox A.~N., Livingston W. C. \& Matthews M.,
Space Science Series, University of Arizona Press, p. 401
\bibref
Christensen-Dalsgaard J. \& Thompson M. J., 1991.
{\it ApJ} {\bf 367}, 666 
\bibref
Christensen-Dalsgaard J. \& Thompson M. J., 1996.
{\it MNRAS}, in the press.
\bibref
Christensen-Dalsgaard J., D\"appen W., Ajukov S. V. {\etal}, 1996.
{\it Science} {\bf 272}, 1286 
\bibref
Christensen-Dalsgaard J., Gough D. O. \& P\'erez Hern\'andez F., 1988.
{\it MNRAS} {\bf 235}, 875 
\bibref
Christensen-Dalsgaard J., Gough D. O. \& Thompson M. J., 1991.
{\it ApJ}
{\bf 378}, 413 
\bibref
Edmonds P., Cram L., Demarque P., Guenther D. B. \&
Pinsonneault M. H., 1992.
{\it ApJ} {\bf 394}, 313 
\bibref
Fernandes J. \& Neuforge C., 1995.
{\it A\&A} {\bf 295}, 678 
\bibref
Gough D. O., 1993.
In {\it Astrophysical fluid dynamics, Les Houches Session XLVII},
eds Zahn J.-P.  \& Zinn-Justin J., Elsevier, Amsterdam,  399 
\bibref
Gough D. O. \& Thompson M. J., 1991.
In {\it Solar interior and atmosphere},
eds Cox A.~N., Livingston W. C. \& Matthews M.,
Space Science Series, University of Arizona Press, p. 519 
\bibref
Gough D. O. \& Toomre J., 1991.
{\it ARA\&A} {\bf 29}, 627 
\bibref
Gough D. O. \& Weiss N. O., 1976.
{\it MNRAS} 
{\bf 176}, 589 
\bibref
Gough D. O., Kosovichev A. G., Toomre J. {\etal}, 1996.
{\it Science} {\bf 272}, 1296 
\bibref
Guzik J. A. \& Cox A.~N., 1995.
{\it ApJ} {\bf 448}, 905 
\bibref
Iglesias C. A., Rogers F. J. \& Wilson B. G., 1992.
{\it ApJ} {\bf 397}, 717 
\bibref
Kosovichev A. G. \& Fedorova A. V., 1991.
{\it Astron. Zh.} {\bf 68}, 1015 
(English translation: {\it Sov. Astron.} {\bf 35}, 507) 
\bibref
Kosovichev A. G., {\etal}, 1992.
{\it MNRAS} {\bf 259}, 536 
\bibref
Michaud G. \& Charbonneau P., 1991.
{\it Space Sci. Rev.} {\bf 57}, 1 
\bibref
Michaud G. \& Proffitt C. R., 1993.
In {\it Proc. IAU Colloq. 137: Inside the stars},
eds Baglin~A. \& Weiss W. W., 
{\it ASP Conf. Ser.} {\bf 40}, 246 
\bibref
Montalb\'an J., 1994.
{\it A\&A} {\bf 281}, 421 
\bibref
Montalb\'an J. \& Schatzman E., 1996.
{\it A\&A} {\bf 305}, 513 
\bibref
Patern\`o L., Ventura R., Canuto V. M. \& Mazzitelli I., 1993.
{\it ApJ} {\bf 402}, 733 
\bibref
Pijpers F. P. \& Thompson M. J., 1992.
{\it A\&A} {\bf 262}, L33 
\bibref
Richard O., Vauclair S., Charbonnel C. \& Dziembowski W. A., 1996.
{\it A\&A} {\bf 312}, 1000 
\bibref
Rogers F. J., Swenson F. J. \& Iglesias C. A., 1996.
{\it ApJ} {\bf 456}, 902 
\bibref
Schwarzschild M., 1958.
{\it Structure and evolution of the stars},
Princeton University Press, Princeton, New Jersey.
\bibref
Stein R. F. \& Nordlund {\AA}., 1989.
{\it ApJ} 
{\bf 342}, L95 
\bibref
Tomczyk S., Schou J. \& Thompson M. J., 1996.
{\it Bull. Astron. Soc. India} {\bf 24}, 245 
\bibref
Tomczyk S., {\etal}, 1995.
{\it Solar Phys.} {\bf 159}, 1 
\bibref
Tripathy S. C., Basu S. \& Christensen-Dalsgaard J., 1996. 
In {\it Sounding Solar and Stellar Interiors. Proc. IAU Symposium No 181,
poster volume},
eds Provost J. \& Schmider F.~X., in the press.


\end{document}